\documentclass{article}
\usepackage[draft]{hyperref} 
\usepackage{setspace}
\usepackage{amsmath}
\usepackage{amssymb}
\usepackage{amsthm}
\usepackage{dsfont}
\usepackage{multirow}
\usepackage{amsfonts}
\usepackage{booktabs}
\usepackage{color}
\usepackage{graphicx}
\usepackage{caption}
\usepackage{tikz}
\usetikzlibrary{fit,calc,positioning,decorations.pathreplacing,matrix}
\usetikzlibrary{shapes,arrows}
\usepackage{subfigure}
\usepackage{caption}

\usepackage[]{algorithmicx}
\usepackage{algpseudocode,algorithm}
\usepackage{cite}
\usepackage{mathtools}
\usepackage{stmaryrd}
\usepackage{multicol,lipsum}
\usepackage[mathscr]{euscript}
\usepackage{mathrsfs}

\usepackage{soul}

\usepackage[normalem]{ulem}
\usepackage{spconf,amsmath,graphicx}

\def\x{{\mathbf x}}

\newcommand{\y}{\mathbf{y}}
\newcommand{\z}{\mathbf{z}}
\newcommand{\s}{\mathbf{s}}
\newcommand{\h}{\mathbf{h}}
\newcommand{\proj}{\mathbf{\Phi}}
\newcommand{\Cs}{\mathbf{C}_{\mathbf{s}}}
\newcommand{\eye}{\mathbf{I}}
\newcommand{\R}{\mathbb{R}}

\title{Unfolding Neural Networks for Compressive Multichannel Blind Deconvolution}
\name{Bahareh Tolooshams$^{\ast\mp}$, Satish~Mulleti$^{\ast\dagger}$, Demba Ba$^{\mp}$, and Yonina C. Eldar$^{\dagger}$
\thanks{$^\ast$ The authors contributed equally to this work.} \thanks{Tolooshams and Ba acknowledge the support of AWS Machine Learning Research Awards. Mulleti and Eldar have received funding from the Benoziyo Endowment Fund for the Advancement of Science, Estate of Olga Klein -– Astrachanthe; European Union’s Horizon 2020 research and innovation program under grant No. 646804-ERC-COG-BNYQ; and the Israel Science Foundation under grant no. 0100101.}}
\address{
$^{\mp}$School of Engineering and Applied Sciences, Harvard University, Cambridge, MA\\
$^{\dagger}$ Faculty of Math and Computer Science, Weizmann Institute of Science, Israel\\
Emails: btolooshams@seas.harvard.edu, satish.mulleti@gmail.com,  demba@seas.harvard.edu,\\ yonina.eldar@weizmann.ac.il}

\begin{document}
\ninept

\maketitle

\begin{abstract}
We propose a learned-structured unfolding neural network for the problem of compressive sparse multichannel blind-deconvolution. In this problem, each channel's measurements are given as convolution of a common source signal and sparse filter. Unlike prior works where the compression is achieved either through random projections or by applying a fixed structured compression matrix, this paper proposes to learn the compression matrix from data. Given the full measurements, the proposed network is trained in an unsupervised fashion to learn the source and estimate sparse filters. Then, given the estimated source, we learn a structured compression operator while optimizing for signal reconstruction and sparse filter recovery. The efficient structure of the compression allows its practical hardware implementation. The proposed neural network is an autoencoder constructed based on an unfolding approach: upon training, the encoder maps the compressed measurements into an estimate of sparse filters using the compression operator and the source, and the linear convolutional decoder reconstructs the full measurements. We demonstrate that our method is superior to classical structured compressive sparse multichannel blind-deconvolution methods in terms of accuracy and speed of sparse filter recovery.

\end{abstract}
\section{Introduction}
\label{sec:intro}
\vspace{-.1in}
In a multi-receiver radar system, a transmit source signal is reflected from sparsely located targets and measured at the receivers. The received signals are modeled as convolutions of the source signal and sparse filters that depend on the targets' location relative to the receivers~\cite{bajwa_radar, bar_radar}. Often, the source signal is unknown at the receiver due to distortion during transmission. With an unknown source, the problem of determining sparse filters is known as sparse multichannel blind-deconvolution (S-MBD). This model is ubiquitous in many other applications such as seismic signal processing~\cite{filho201seismic}, room impulse response modeling~\cite{rip}, sonar imaging~\cite{carter_sonar}, and ultrasound imaging~\cite{eldar_sos,eldar_beamforming}. In these applications, the receivers' hardware and computational complexity depend on the number of measurements required at each receiver or channel to determine sparse filters uniquely. Hence, it is desirable to compress the number of measurements on each channel.

Prior works have proposed computationally efficient and robust algorithms to recover the filters from the measurements. Examples are $\ell_1$-norm methods~\cite{wang_chi, kazemi2014sparse, bilen}, the sparse dictionary calibration~\cite{gribonval_dl, mulleti_mbd} and truncated power iteration methods~\cite{li2017blind}, and convolutional dictionary learning~\cite{garcia_cdl}. The works in~\cite{lee_17, wang_chi, cosse_mbd} establish theoretical guarantees on identifying the S-MBD problem. However, these methods are computationally demanding, require a series of iterations for convergence, and need access to the full measurements.

Recent works proposed model-based neural networks to address computational efficiency \cite{tolooshams2020tnnls, tolooshams2018mlsp}, but still require full measurements for recovery. To enable compression, Chang et al. \cite{chang2019randnet} proposed an autoencoder, called RandNet, for dictionary learning. In this work, compression is achieved by projecting data into a lower-dimensional space through a data-independent unstructured random matrix. Mulleti et al.~\cite{mulleti_mbd} proposed a data-independent, structured compression operator that could be realized through a linear filter. Specifically, the compressed measurements are computed by applying a specific filter to the received signals followed by the truncation of the filtered measurements. Two natural questions are (1) can we learn a compression filter from a given set of measurements, rather than applying a fixed filter as in~\cite{mulleti_mbd}? (2) will this data-driven approach result in better compression for a given estimation accuracy?

We propose a model-based neural network that learns a hardware-efficient, data-driven, and structured compression matrix to recover sparse filters from reduced measurements. Our approach takes inspiration from filter-based compression~\cite{mulleti_mbd}, learning compression operators~\cite{mousavi}, and the model-based compressed learning approach for S-MBD~\cite{chang2019randnet}. The architecture, which we call learned (L) structured (S) compressive multichannel blind-deconvolution network (LS-MBD), learns a filter for compression, recovers the unknown source, and estimates sparse filters. In contrast to~\cite{mulleti_mbd, mousavi, chang2019randnet}, LS-MBD improves in the following ways. In \cite{mulleti_mbd}, a fixed and data-independent filter is used for compression, and the reconstruction is independent of the compression. In LS-MBD, we learn a filter that enables compression \emph{and} lets us estimate sparse filters accurately. The approach is computationally efficient and results in lower reconstruction error for a given compression ratio compared with \cite{mulleti_mbd}. Unlike in~\cite{mousavi}, our compression operator is linear, is used recurrently in the architecture, has fewer parameters to learn, and has computationally efficient implementation~\cite{durate_eldar}.

Section~\ref{sec:pf} explains the S-MBD problem formulation. In Section~\ref{sec:network}, we introduce our method, its architecture and training procedure. We present our results in Section~\ref{sec:exp}, highlighting the superiority of LS-MBD and its efficiency compared to baselines.
\begin{figure*}[!t]
	\begin{minipage}[b]{1.0\linewidth}
		\centering
		\centering
		\tikzstyle{block} = [draw, fill=none, rectangle, 
		minimum height=2em, minimum width=2em]
		\tikzstyle{sum} = [draw, fill=none, minimum height=0.1em, minimum width=0.1em, circle, node distance=1cm]
		\tikzstyle{cir} = [draw, fill=none, circle, line width=0.7mm, minimum width=0.65cm, node distance=1cm]
		\tikzstyle{loss} = [draw, fill=none, color=black, ellipse, line width=0.5mm, minimum width=0.7cm, node distance=1cm]
		\tikzstyle{blueloss} = [draw, fill=none, color=blue!70, ellipse, line width=0.3mm, minimum width=0.5cm, node distance=1cm]
		\tikzstyle{orangeloss} = [draw, fill=none, color=orange, ellipse, line width=0.3mm, minimum width=0.5cm, node distance=1cm]
		\tikzstyle{relublock} = [draw, rectangle, fill=none,
		minimum height=2em, minimum width=2em]
		\tikzstyle{orangeblock} = [draw, rectangle, fill=orange, color=orange, minimum height=2em, minimum width=2em]
		\tikzstyle{blueblock} = [draw, rectangle, fill=blue, color=blue!70, minimum height=2em, minimum width=2em]
		\tikzstyle{input} = [coordinate]
		\tikzstyle{output} = [coordinate]
		\tikzstyle{pinstyle} = [pin edge={to-,thin,black}]
		\begin{tikzpicture}[auto, node distance=2cm,>=latex']
		cloud/.style={
			draw=red,
			thick,
			ellipse,
			fill=none,
			minimum height=1em}
		\node [input, name=input] {};
		\node [cir, node distance=0cm, right of=input] (Y) {$\y$};
		\node [blueblock, right of=Y,  minimum width=0.5cm, node distance=1.23cm] (PHI) {$\mathbf{\Phi}$};
		\node [block, right of=Y,  minimum width=0.5cm, node distance=1.23cm] (PHI) {$\mathbf{\Phi}$};
		\node [cir, node distance=1.cm, right of=PHI] (r) {$\z$};

		\node [block, right of=r,  minimum width=0.5cm, node distance=1.2cm] (alpha) {$\alpha$};
		\node [blueblock, right of=alpha,  minimum width=0.5cm, node distance=1.0cm] (PHIT) {$\mathbf{\proj^{\text{T}}}$};
		\node [block, right of=alpha,  minimum width=0.5cm, node distance=1.0cm] (PHIT) {$\mathbf{\proj^{\text{T}}}$};
		\node [orangeblock, right of=PHIT,  minimum width=0.5cm, node distance=1.2cm] (ST) {$\Cs^{\text{T}}$};
		\node [block, right of=PHIT,  minimum width=0.5cm, node distance=1.2cm] (ST) {$\Cs^{\text{T}}$};
		
		\node [relublock, right of=ST,  minimum width=0.7cm, node distance=2cm] (relu) {$\mathcal{S}_{b_t}$};
		
		\node [block, below of=relu,  minimum width=0.7cm, node distance=0.8cm] (f) {$f_{\textcolor{orange}{\Cs}, \textcolor{blue!70}{\proj}}(.)$};
		
		\node [cir, right of=relu, node distance=1.2cm] (xt) {$\x_{t}$};
		
		\node [cir, right of=xt, node distance=1.2cm] (xT) {$\hat \x$};
		\node [cir, below of=xT, node distance=1cm] (xtrain) {$\tilde \x$};
	
		\node [blueloss, node distance=1.8cm, right of=xtrain] (loss-en) {$\| \tilde \x - \hat \x \|_2^2$};	
		\node [rectangle, fill=none,  node distance=0.57cm,  left=2.7pt,above of=loss-en] (loss-en-des) {\footnotesize{Stage 2 Loss}}; 	

		\node [orangeblock, right of=xT, node distance=2.5cm] (S) {};
		\node [block, right of=xT, node distance=2.5cm] (S) {$\Cs$};
		\node [cir, node distance=1.3cm, right of=S] (Yhat) {$\hat \y$};
		\node [cir, node distance=1.cm, below of=Yhat] (Ytrain) {$\y$};
	
		\node [orangeloss, node distance=1.8cm, right of=Ytrain] (loss-ae) {$\| \y - \hat \y \|_2^2$};	
		\node [rectangle, fill=none,  node distance=0.57cm,  left=2.7pt,above of=loss-ae] (loss-des) {\footnotesize{Stage 1 Loss}}; 	
				
		\draw[thick, line width=2, black, ->]     ($(xt.north east)+(1.2,0.48)$) -- ($(xt.north east)+(5,0.48)$);
		\draw[thick, line width=2, black, ->]     ($(xt.north east)+(-6,0.48)$) -- ($(xt.north east)+(1.,0.48)$);
		
   		\node [rectangle, fill=none,  node distance=0.565cm,  above of=relu] (text) {\footnotesize{Repeat $T$ times}};
		\node [rectangle, fill=none,  node distance=0.92cm,above of=S] (text) {\footnotesize{Decoder}};
		\node [rectangle, fill=none,  node distance=0.92cm,  above of=relu] (encoder) {\footnotesize{Encoder}}; 
		
		\draw[thick,dotted]     ($(xt.north east)+(+0.2,+0.17)$) rectangle ($(relu.west)+(-0.98,-1.2)$);
		
		\draw[thick,dotted]     ($(PHI.east)+(+0.15,+0.4)$) rectangle ($(Y.west)+(-0.15,-0.4)$);
		
		\node [rectangle, fill=none,  node distance=0.82cm,  left=17.0pt,above of=PHI] (training) {\footnotesize{Available only}}; 
		\node [rectangle, fill=none,  node distance=0.57cm,  left=17.0pt,above of=PHI] (training) {\footnotesize{for Training}};

		\draw [->] (Y) -- node [] {} (PHI);
		\draw [->] (PHI) -- node [] {} (r);		
		\draw [->] (r) -- node [] {} (alpha);
		\draw [->] (alpha) -- node {} (PHIT);
		\draw [->] (PHIT) -- node {} (ST);
		\draw [->] (ST) -- node[name=s, pos=0.4, above] {} (relu);
		\draw [->] (relu) -- node[] {} (xt);
		\draw [-] (xt) |- node[] {} (f);
		\draw [->] (f) -| node[] {} (s);
		\draw [->] (xt) -- node[] {} (xT);
		\draw [->] (xT) -- node[] {} (S);	
		\draw [->] (S) -- node[] {} (Yhat);	
	
		\draw [dashed] (xT) -- node[] {} (loss-en);	
		\draw [dashed] (xtrain) -- node[] {} (loss-en);		
		\draw [dashed] (Yhat) -- node[] {} (loss-ae);	
		\draw [dashed] (Ytrain) -- node[] {} (loss-ae);	
		\end{tikzpicture}
	\end{minipage}
	\vspace{-8mm}
	\caption{Architecture of Learned Structured Multichannel Blind-Deconvolution (LS-MBD). $f_{\Cs, \proj}(\x_t, \x_{t-1}) = (\eye - \alpha \Cs^{\text{T}} \proj^{\text{T}} \proj \Cs) (\x_t + \frac{s_t - 1}{s_{t+1}} (\x_t - \x_{t-1}))$, where we set $s_t$ as in FISTA~\cite{beck2009fast}. In the first stage of training, we set $\proj = \eye$ to have access to the full measurements. Trainable blocks/weights and the loss for the first and second stage of training are in orange and blue, respectively.}\label{fig:ls-mbd}
\vspace{-5mm}
\end{figure*}
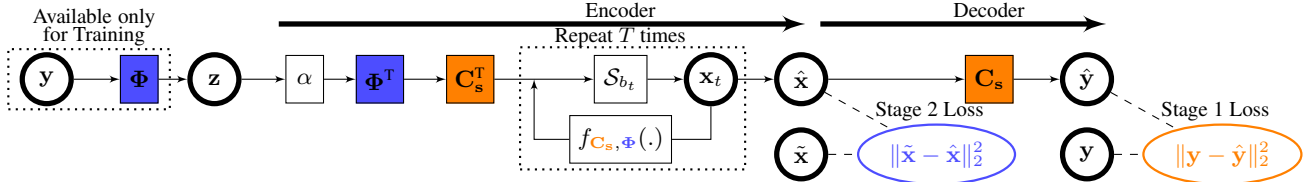

\vspace{-3mm}
\section{Problem Formulation}
\label{sec:pf}
\vspace{-3mm}
Consider a set of $N$ signals given as
\vspace{-3mm}
\begin{align}
    \y^n = \mathbf{s}*\mathbf{x}^n = \Cs\mathbf{x}^n, \quad n =1,\ldots, N,
    \label{eq:mbd1}
\end{align}
where $*$ denotes the linear-convolution operation and the matrix $\mathbf{C}_{\s}$ is the convolution matrix constructed from the vector $\s$. We make the following structural assumption on the signals: (A1) $\s\in \R^{M_s}$ and (A2) $\x^n\in \R^{M_x}$ and $\|\mathbf{x}^n\|_0\leq L$ for $n = 1, \ldots, N$.
Due to convolution, each $\y^n$ has length $M_y = M_x+M_s-1$. In a S-MBD problem, the objective is to determine the common source signal $\mathbf{s}$ and sparse filters $\mathbf{X} = [\mathbf{x}^1, \mathbf{x}^2, \cdots, \mathbf{x}^N] \in \mathbb{R}^{M_x \times N}$ from the measurements $\mathbf{Y} = [\y^1, \y^2, \cdots, \y^N] \in \mathbb{R}^{M_y \times N}$. 

In compressive S-MBD, the objective is to estimate the unknowns from a set of compressive measurements $\mathbf{z}^n = \boldsymbol{\Phi}(\y^n) \in \mathbb{C}^{M_z}, n = 1, \ldots, N$ where $M_z < M_y$. The compression operator $\boldsymbol{\Phi}$ is a mapping from $\R^{M_y}$ to $\mathbb{C}^{M_z}$. This operator could be either linear or nonlinear, random or deterministic, structured or unstructured, data-dependent or data-independent (see \cite{durate_eldar} for details). 

We consider the problem of designing a linear, structured, and data-driven compression operator that enables accurate estimation of sparse filters from the compressed measurements. Specifically, our goal is to jointly learn $\s$, and a structured, practically-realizable and data-driven $\proj$ such that sparse filters, $\mathbf{X}$, are determined from the compressive measurements $\{\z^n = \proj\y^n\}_{n=1}^N$.


\vspace{-5mm}
\section{Learning Structured Compressive S-MBD}\label{sec:network}
\subsection{Compression Operator}
We impose a Toeplitz structure on the compression operator $\proj$ and denote the filter associated with the operator by $\h$. Such compression matrices have several advantages compared to random projections; they facilitate the use of computationally efficient matrix operations and can be practically implemented as a filtering operation in hardware. The compression operator involves a convolution followed by a truncation. As the operation holds for all $n$, we drop the channel-index superscript, for simplicity. Consider a causal filter $\h$ of length $M_h>M_y$. The truncated convolution samples are
\vspace{-2mm}
\begin{align}
     (\y*\h)[m] = \sum_{l=0}^{M_y-1} \y[l]\h[m-l], \,\, M_y-1 \leq m \leq M_h-1,
         \label{eq:conv}
\end{align}
which are the convolution samples where $\h$ and $\y$ have complete overlap.
Choice of samples $M_y-1 \leq m \leq M_h-1$ is to ensure that maximum information from $\y$ is retained in $\z$.
The corresponding measurement matrix $\boldsymbol{\Phi}$ is a $M_z \times M_y$ Toeplitz matrix given as
\begin{align}
    \proj \hspace{-.03in} = \hspace{-.06in}
    \begin{pmatrix}
\h[M_y-1] & \h[M_y-2] & \cdots & \h[0] \\
\h[M_y] & \h[M_y-1] & \cdots & \h[1] \\
\h[M_y+1] & \h[M_y] &  \cdots & \h[2] \\
\vdots  & \vdots &  \ddots & \vdots  \\
\h[M_y+M_z-2] & \h[M_y+M_z-3] &  \cdots & \h[M_z-1] \\ 
    \end{pmatrix}\hspace{-.04in}.\nonumber
    \label{eq:h}
\end{align}

\subsection{Network Architecture}
We aim to minimize the following objective
\vspace{-3mm}
\begin{equation}\label{eq:obj}
\begin{aligned}
\min_{\proj, \s, \{\x^n\}_{n=1}^N}&\ \frac{1}{2} \sum_{n=1}^N \| \z^n - \proj \Cs \x^n \|_2^2 + \lambda \| \x^n \|_1\\
&\text{s.t.}\ \|\s\|_2 = \| \h \|_2 = 1
\end{aligned}
\end{equation}
where $\lambda$ is a sparsity-enforcing parameter, and the norm constraints are to avoid scaling ambiguity. Following a similar approach to~\cite{gregor2010learning, tolooshams2020tnnls, monga2019algorithm}, we construct an autoencoder where its encoder maps $\z^n$ into a sparse filter $\x^n$ by unfolding $T$ iterations of a variant of accelerated proximal gradient algorithm, FISTA~\cite{beck2009fast}, for sparse recovery. Specifically, each unfolding layer performs the following iteration
\begin{equation}\label{eq:enc}
\x_t^n = \mathcal{S}_{b_t} (f_{\Cs, \proj}(\x_{t-1}^n, \x_{t-2}^n) + \alpha \Cs^{\text{T}} \proj^{\text{T}} \z^n) 
\end{equation}
where $f_{\Cs, \proj}(\x_t, \x_{t-1}) = (\eye - \alpha \Cs^{\text{T}} \proj^{\text{T}} \proj \Cs) (\x_t + \frac{s_t - 1}{s_{t+1}} (\x_t - \x_{t-1}))$, $s_t$ is set as in FISTA, $\alpha$ is the unfolding step-size, and the sparsity promoting soft-thresholding operator $\mathcal{S}_{b}(v) = \text{ReLU}_b(v) - \text{ReLU}_b(-v)$ where $\text{ReLU}_b(v) = (v - b) \cdot \mathds{1}_{v \geq b}$.

One may leave the bias $b_t$ in each layer unconstrained to be learned. However, theoretical analysis of unfolding ISTA~\cite{chen2018theoretical} has proved that $b_t$ converges to $0$ as $t$ goes to infinity. Hence, we set
$b_t = \alpha \lambda_t = \alpha c^t \lambda$ with $c \in (0, 1]$. In this regard, $\lambda$ is a scalar that can be either tuned or learned. We keep the unfolded layers tied as this leads to better generalization in applications with limited-data.

The decoder reconstructs the data $\hat{\y}^n$ using $\Cs$. We call this network, shown in Figure~\ref{fig:ls-mbd}, LS-MBD. In this architecture, $\proj$ and $\Cs$ correspond to weights of the neural network and are learned by backpropagation in two stages. This method's novelty is in the hardware-efficient design of the operator $\proj$ capturing data information when compressing. This architecture reduces to RandNet~\cite{chang2019randnet} when $\proj$ and $\Cs$ have no structure (e.g. Toeplitz) and $c=1$. 

\vspace{-3mm}
\subsection{Training Procedure}
We follow a two-stage approach to learn the compression filter. In the first stage, we start with a limited set of full measurements and estimate the source and filters. In the second, we learn the compression filter associated with $\proj$ given the estimated source. This two-stage training disentangles source estimation and learning of the compression. Hence, source estimation is performed only once to be used for various compression ratios (CR)s. Besides, a low CR would not affect the quality of source estimation, and the number of measurements required for filter recovery can be optimized independently.

Specifically, having access to the full measurements $\y$, we set the compression operator to be the identity matrix (i.e., $\proj = \eye$), hence reducing the autoencoder architecture to a variant of CRsAE~\cite{tolooshams2018mlsp}, and learn $\Cs$ by minimizing $\| \y - \hat \y \|_2^2$. Then, given the learned source matrix $\Cs$, we run the forward set of the encoder to estimate sparse filters, which we denote $\tilde \x $. In the second stage, for a given compression ratio $\text{CR} = M_z/M_y$, we set $\Cs$ to the learned source and train $\proj$ within the encoder by minimizing $\| \hat \x -  \tilde \x\|_2^2$.

In practice, the method is useful in the following scenario. Consider a multi-receiver radar scenario where a set of receivers are already in place and operate with full measurements. Let it require adding a new set of receivers to replace the existing ones or gather more information. While designing these new set of receivers, one can use the information from the available full measurements from the existing receivers to estimate the source and then design the optimal compression filters for new receivers. Thus, the new receivers sense compressed measurements that result in reduced cost and complexity. In essence, the approach is similar to the deep-sparse array method~\cite{deepLearningAntennaSelectionElbir}, where a full array is used in one scan to learn sparse arrays for the rest of the scans.

\vspace{-4mm}
\section{Experiments}
\label{sec:exp}

\vspace{-2mm}
\subsection{Data Generation}
We considered the noiseless case and generated $N=100$ measurements following the model in \eqref{eq:mbd1} where $M_s = 99$, $M_x = 100$, and $L=6$. The source follows a Gaussian shape generated as $\s[k] = e^{-6 (k - \lfloor\frac{M_s}{2}\rfloor)^2}$ where $k \in \{0, 2, \ldots, M_s-1\}$, and then normalized such that $\| \s \|_2 = 1$. The support of the filters are generated uniformly at random over the set the set $\{0, \ldots, M_x-1\}$, and their amplitudes from a uniform distribution $\text{Unif}(0,1)$.

In the aforementioned assumptions, we neither impose restrictions on the minimum separation of any two non-zero components of the sparse filter nor the the filter components' relative amplitudes. In the presence of noise, both of these factors play a crucial role in the filter estimation, and the recovery error is expected to increase. Our method's recovery performance and stability analysis in the presence of noise is a direction of future research.


\vspace{-4mm}
\subsection{Network and Training Parameters}

We implemented the network in PyTorch. We unfolded the encoder for $T= 15{,}000$ iterations, and set $\alpha = 0.05$. We set the regularization parameter $\lambda = 0.1$ and decreased by a factor of $c = 0.99937$ at every iteration (i.e., $\lambda_t = c^t \lambda$). To ensure stability, we chose the parameter $\alpha$ to be less than the reciprocal of the singular value of the matrix $\proj \Cs$ at initialization. In the second stage, we tuned $\lambda$ and $c$ finely by grid-search in the ranges $[0.1, 0.5]$ and $[0.98,1]$, respectively. 
Given the non-negativity of sparse filters, we let $\mathcal{S} = \text{ReLU}$.

In the first stage, we initialize the source's entries randomly according to a standard Normal distribution. The network is trained for $1{,}000$ epochs using gradient descent. We use the ADAM optimizer with a learning rate of $0.03$, decreased by a factor of $0.9$ every $100$ epochs. To achieve convergence without frequent skipping of the minimum of interest, we set $\epsilon$ of the optimizer to be $10^{-2}$. Given the learned source, and following a similar approach in generating sparse filters, we generate $10{,}000$ examples for training, $100$ for validation, and $100$ for testing. We run the encoder given $\y$ to estimate sparse filters $\tilde \x$. In the second stage, we initialize the filter for $\proj$ similarly to the source. We use $\epsilon = 10^{-8}$ and a learning rate of $10^{-3}$ with a decaying factor of $0.9$ for every $100$ epochs. We use a batch size of $100$ and train for $1{,}000$ epochs for each CR.

The iterative usage of $\proj$ and $\Cs$ within each layer of the architecture, a property that does not exist in~\cite{mousavi}, allows us to use a combination of analytic and automatic differentiation~\cite{ablin2020super} to compute the gradient; in this regard, we backpropagate up to the last $100$ iterations within the encoder to compute the gradient and assume that the representation $\x_{T-100}$ is not a function of the trainable parameters. This allows us to unfold the network for large $T$ in the forward pass without increasing the computational and space complexity of backpropagation, a property that unfolded networks with untied layers do not possess. Lastly, we normalize the source and compression filter after every gradient update.

\vspace{-4mm}
\subsection{Baselines}

For each CR, we compare five methods detailed below.

\noindent \textbf{LS-MBD}:\quad $\proj$ is learned (L) and structured (S).

\noindent \textbf{LS-MBD-L:}\quad $\proj$ is learned (L) and structured (S). Motivated by the fast convergence of LISTA, the encoder performs $T=20$ steps of the proximal gradient iteration $\x_t^n = \mathcal{S}_{b} ((\eye - \mathbf{W}_e \proj^{\text{T}} \proj \mathbf{W}_d)\x_{t-1}^n + \mathbf{P} \proj^{\text{T}} \z^n)$. In the second stage of training, we learn the bias $b$, and convolution/correlations operators $\mathbf{W}_d,  \mathbf{W}_e, \mathbf{P}, \proj$.

\noindent \textbf{GS-MBD:}\quad $\proj$ is random Gaussian (G) and structured (S).

\noindent \textbf{FS-MBD:}\quad $\proj$ is fixed (F) and structured (S)~\cite{mulleti_mbd}. In~\cite{mulleti_mbd}, the authors derive identifiability in the Fourier domain, and  design the filters to enable computation of the specific Fourier measurements. The authors also show that for FS-MBD, the $L$-sparse filters are uniquely identifiable from $2L^2$ compressed measurements from any two channels and $2L$ compressed measurements from the rest of the channels. Here, we applied the blind-dictionary calibration approach from~\cite{mulleti_mbd} together with FISTA~\cite{beck2009fast} for the sparse-coding step.


\noindent \textbf{G-MBD:}\quad $\proj$ is an unstructured random Gaussian (G) matrix. We consider G-MBD as an oracle baseline that implements a computationally expensive compression operator.

LS-MBD, GS-MBD, and G-MBD use the architecture shown in Figure~\ref{fig:ls-mbd}, each with a different $\proj$. 

\vspace{-4mm}
\subsection{Results}

We show that LS-MBD is superior to GS-MBD, FS-MBD, and LS-MBD-L. We evaluate performance in terms of how well we can estimate the filters and source. Let $\hat{\mathbf{X}}$ be an estimate true filters $\mathbf{X}$. We use the normalized MSE $20 \log{(\| \mathbf{X} - \hat{\mathbf{X}}\|_2 / \| \mathbf{X} \|_2)}$ in dB as a comparison metric and call a method successful if this error is below $-50$ dB. Letting $\hat{\s}$ denote an estimate of the source, we quantify the quality of source recovery using the error $\sqrt{ 1 - \langle \s, \hat{\s}\rangle}$~\cite{Agarwal2016LearningSU}, which ranges from $0$ to $1$, where $0$ corresponds to exact source recovery.

\begin{figure}[!t]
\begin{center}
\begin{tabular}{cc}
\subfigure[]{\includegraphics[width= 1.6 in]{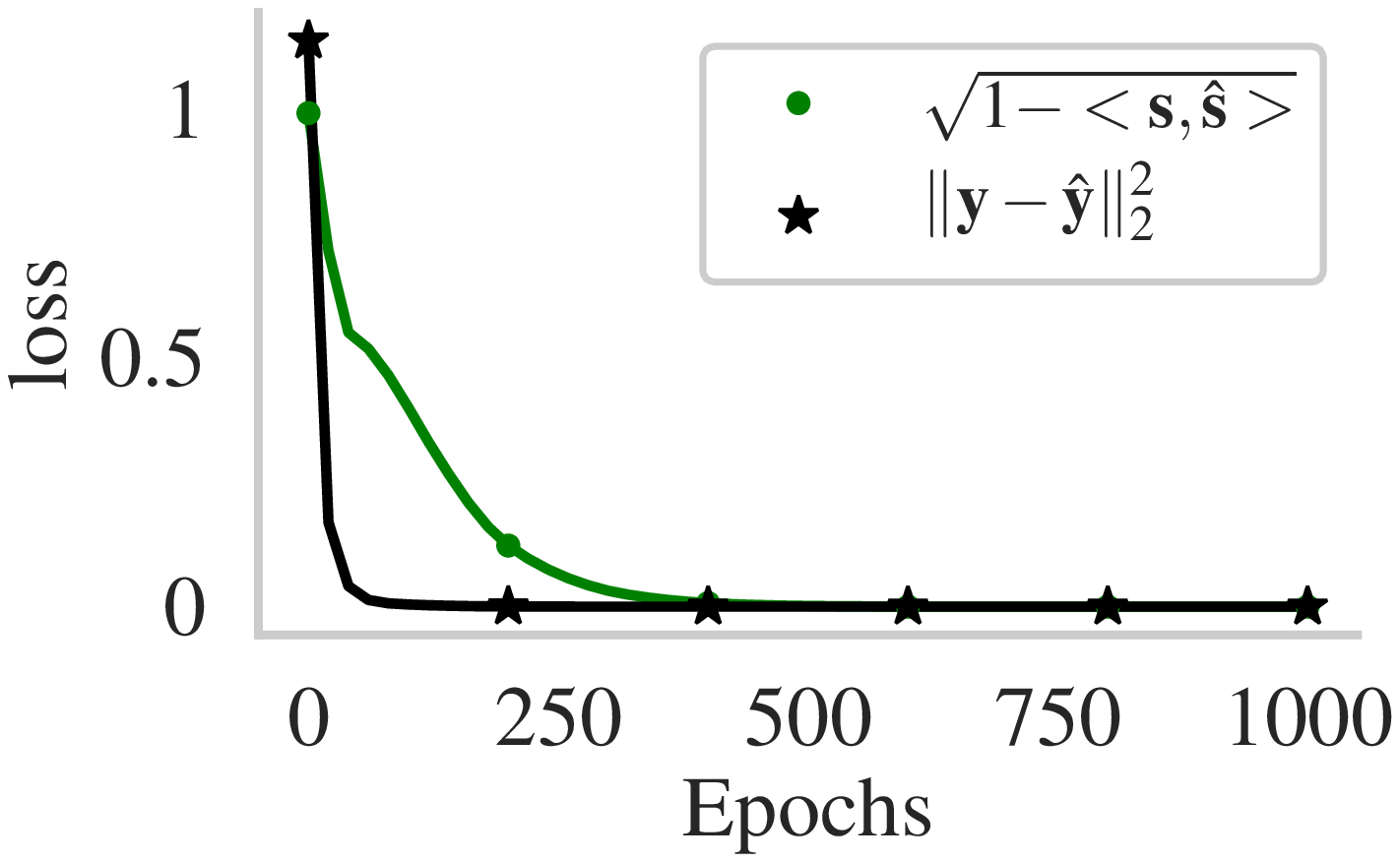}\label{fig:stage1-loss}} 
\subfigure[]{\includegraphics[width= 1.6 in]{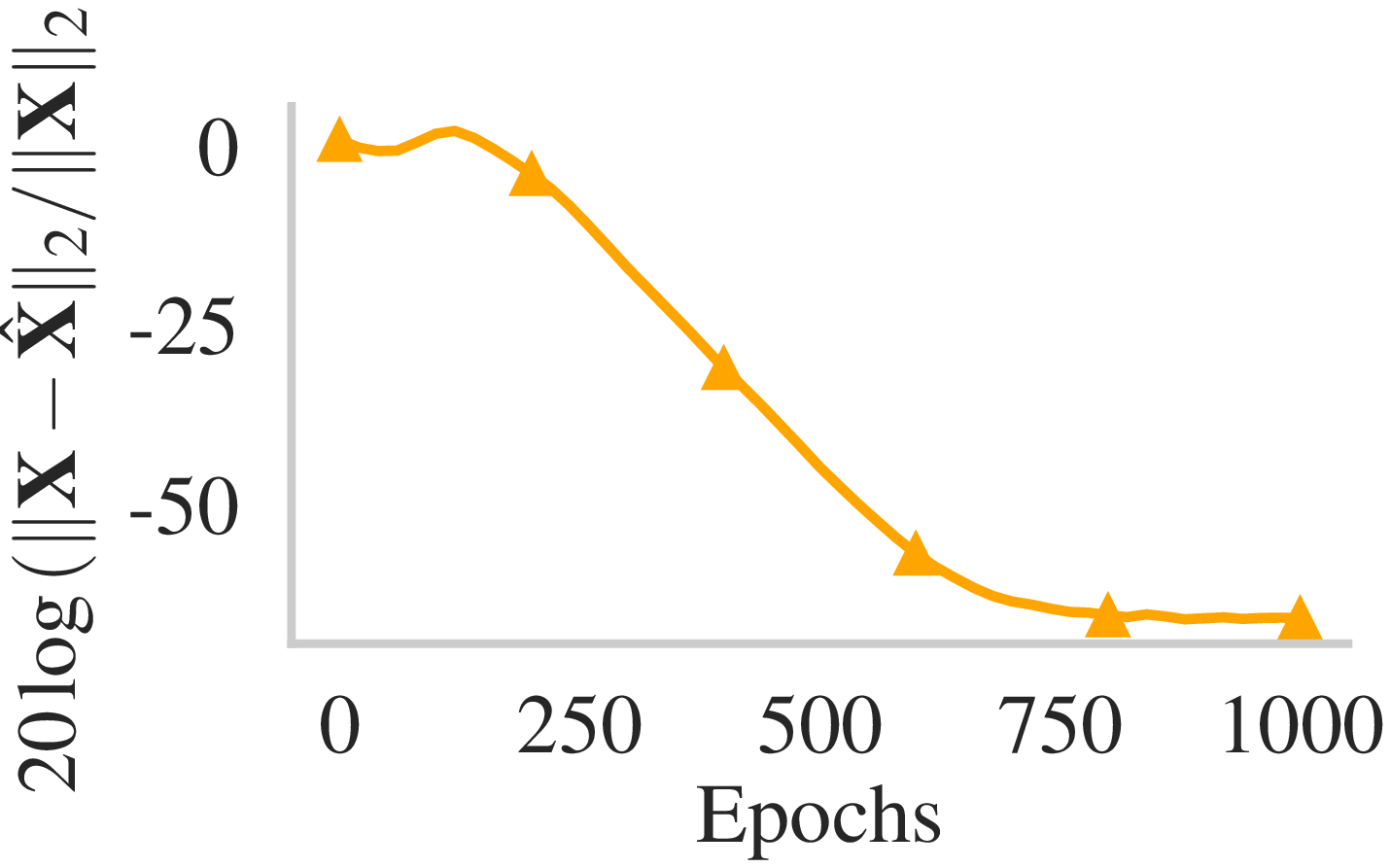}\label{fig:stage1-filter-loss}}\\
\subfigure[]{\includegraphics[width=1.6in]{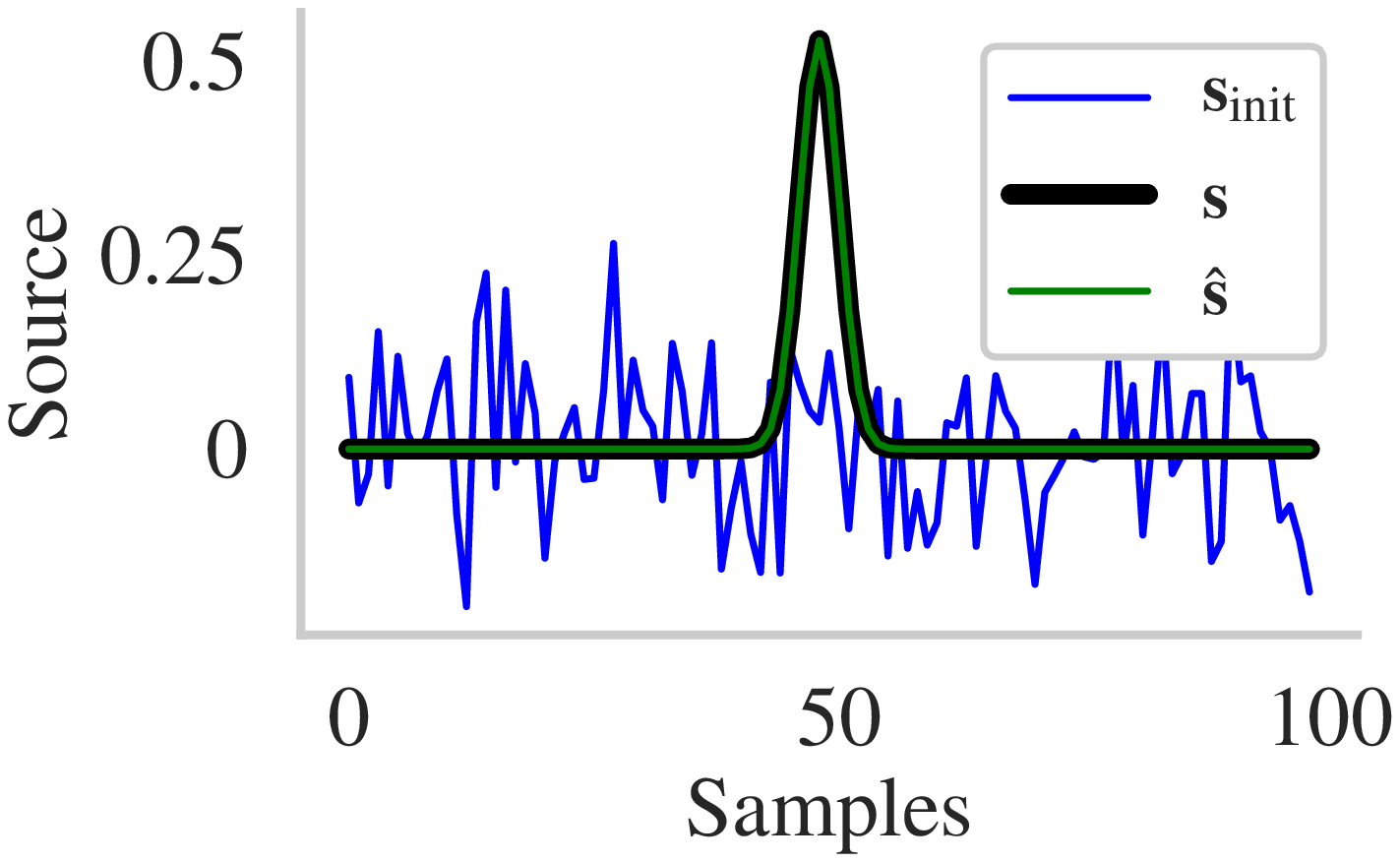}\label{fig:stage1-s}}
\subfigure[]{\includegraphics[width=1.6in]{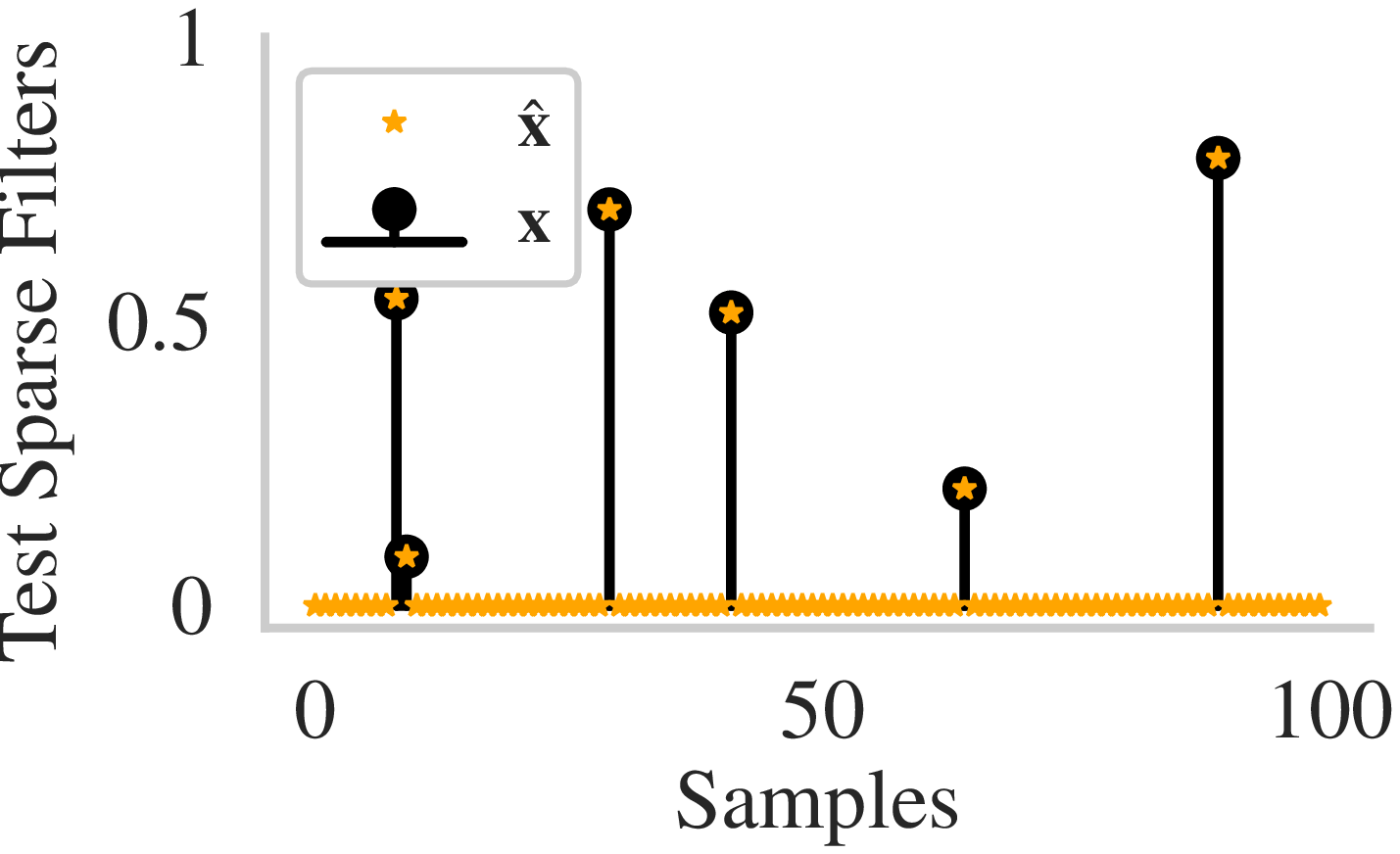}\label{fig:stage2-x}}
\end{tabular}
\vspace{-6mm}
\caption{LS-MBD: (a) Reconstruction and source estimation losses as a function of epochs during the first stage of training. (b) Sparse filter recovery error as a function of epochs during the first stage of training. (c) Initialized (blue), learned (green), and true (black) source. (d) Sparse filter recovery of a test example for $\text{CR}=23.74\%$.}
\label{fig:dirac_estimation}
\end{center}
\vspace{-8mm}
\end{figure}

Figures~\ref{fig:stage1-loss},~\ref{fig:stage1-filter-loss}, and~\ref{fig:stage1-s} visualize the LS-MBD results from the first stage of training. Figure~\ref{fig:stage1-loss} shows that the source estimation error and the training loss both decrease and converge to zero as a function of epochs. Figure~\ref{fig:stage1-filter-loss} shows the filter estimation error and successful recovery of the filters upon training. Figure~\ref{fig:stage1-s} shows the source before and after training, where the learned and true sources match. Figure~\ref{fig:stage2-x} visualizes the filter recovered using a test example obtained with a CR of $23.74\%$ after full training.

Table~\ref{tab:runtime} shows the inference runtime of the methods averaged over 20 independent trials. All methods except FS-MBD are run on GPU. LS-MBD-L has the fastest inference because it only unfolds a small number of proximal gradient iterations. Despite its speed, LS-MBD-L has the worst recovery performance.

\begin{table}[!t]
 \renewcommand{\arraystretch}{0.9}
  \centering
  \caption{Comparison of the inference runtime of the methods for when $\text{CR}=50\%$.}
  \setlength\tabcolsep{4pt}
  \begin{tabular}{c|cccc}
    \bottomrule
   Speed $\backslash$ Method & G-MBD & FS-MBD & LS-MBD & LS-MBD-L\\
     \hline \hline
    runtime [s] & $4.9087$ & 164 & $5.4204$ & ${\bf 0.0028}$\\
    \hline
    \bottomrule
  \end{tabular}
  \vspace{-4mm}
  \label{tab:runtime}
\end{table}
\begin{table}[!t]
  \renewcommand{\arraystretch}{0.9}
  \centering
  \caption{Filter recovery error in dB on the test set for various CRs (bold indicates successful recovery with error below $-50$ dB).}
  \setlength\tabcolsep{2pt}
  \begin{tabular}{cc|ccccc}
    \bottomrule
    CR [\%] & $M_z$ & G-MBD & GS-MBD & FS-MBD & LS-MBD & LS-MBD-L\\
     \hline \hline
    $50$ & 99 & {\bf -54.05} & -44.93 & -43.96 & {\bf -53.27} & -26.54\\ 
    $40.4$ & 80 & {\bf -55.07} & -40.55 & -26.52 & {\bf -52.80} & -\\ 
    $35.35$ & 70 & {\bf -52.43} & -40.00 &  -22.76 & {\bf -51.50} & -\\ 
    $31.31$ & 62 & {\bf -53.63} & -37.13 & -21.86 & {\bf -54.71} & -\\ 
    $25.25$ & 50 & {\bf -53.36} & -28.57 & -8.40 & {\bf -51.41} & -\\ 
    $23.74$ & 47& {\bf -50.60} & -26.11 & -6.84 & {\bf -50.35} & -\\ 
    $22.72$ & 45 & {\bf -52.98} & -23.17 & -6.14 & -43.61 & -\\ 
    $20.20$ & 40 & -47.39 & -14.75 & -5.13 & -17.07 & -\\ 
    \hline
    \bottomrule
  \end{tabular}
  \label{tab:res}
  \vspace{-4mm}
\end{table}

Table~\ref{tab:res} shows the filter recovery error for various CR on the test set. G-MBD and GS-MBD results are averages over ten independent trials. For LS-MBD-L, we only report for $\text{CR} = 50\%$, which already shows a very high recovery error. Among the three structured compression matrix methods, we observe that the proposed LS-MBD approach outperforms other structured methods. Theoretical results in~\cite{mulleti_mbd} suggest that for structured compression when $M_z\geq 2L^2 = 72$ (i.e., $\text{CR} \geq 36.36\%$), we will have successful recovery of sparse filters. LS-MBD goes beyond and has a successful recovery for CR lower than $36.36\%$ up to $23.74\%$.

Comparison of LS-MBD and LS-MBD-L highlights the importance of deep unfolding/encoder depth and the use of a model-based encoder (i.e., large $T$ and tied encoder/decoders). LS-MBD-L, even in the presence of a learned encoder, fails to recover sparse filters. Besides, we observed that LS-MBD generalizes better (i.e., closer test and training errors) than LS-MBD-L, highlighting the importance of limiting the number of trainable parameters by weight-tying in applications where data are limited.

For FS-MBD, we observe that for the given set of data, the approach fails to estimate the filters for $M_z\geq 2L^2$ accurately. In large part, this is due to the failure of the FISTA algorithm in the sparse-coding step. To verify, we consider the non-blind case where the source is assumed to be known. In this case, the normalized MSE in the estimation of the filters from the compressed Fourier measurements by using FISTA are comparable to that of FS-MBD. In comparison to \cite{mulleti_mbd}, LS-MBD learns  a filter  that  enables  compression and accurate sparse coding, which results in a lower MSE. We observe that our oracle method, G-MBD, outperforms the structure-based compression methods. We attribute this superior performance to the fact that the compression matrix in G-MBD has independent random entries that result in low mutual coherence among its columns \cite{eldar_cs_book}. In the rest of the methods, the compression matrices have a Toeplitz structure, which results in high coherence. Despite being less accurate compared to G-MBD, the compression matrix in LS-MBD has fewer degrees of freedom, is practically feasible to implement, and its Toeplitz structure can be used to speed up matrix computations in the recovery process. Table~\ref{tab:comp} compares the memory storage and computational costs of the unstructured (G-MBD) and structured (LS-MBD) compression operators. The table highlights the efficiency of LS-MBD; in this case, we report the complexity of the operation performed using the fast Fourier transform.
 
 \begin{table}[htb]
 \vspace{-5mm}
   \renewcommand{\arraystretch}{0.9}
  \centering
  \caption{Memory storage and computational complexity of the compression operator $\proj$ when it is structured or unstructured.}
  \setlength\tabcolsep{11pt}
  \begin{tabular}{c|cc}
    \bottomrule
    Method $\backslash$ Cost &  Memory Storage & Complexity\\
     \hline \hline
    Structured & ${ O(M_h)}$ & ${O(M_h\log{M_h})}$\\
    Unstructured & $O(M_yM_z)$ & $O(M_yM_z)$\\
    \hline
    \bottomrule
  \end{tabular}
  \label{tab:comp}
 \vspace{-2mm}
\end{table}

\begin{figure}[!t]
$$\includegraphics[width=0.7\linewidth]{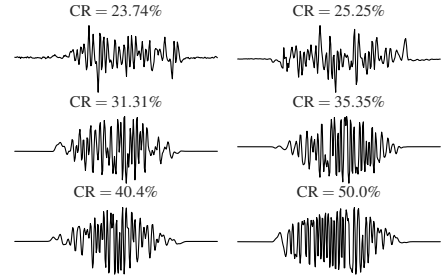}$$
\vspace{-10mm}
\caption{Learned compression filter corresponding to compression matrix $\proj$ in LS-MBD for various CR shown in time domain.}\label{fig:proj}
 \vspace{-3mm}
\end{figure}

\begin{figure}[!t]
$$\includegraphics[width=0.7\linewidth]{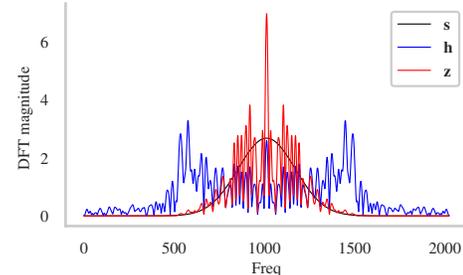}$$
\vspace{-10mm}
 \caption{Magnitude of discrete Fourier transform of the source (black), the learned compression filter (blue), and the compressed measurements (red) for $\text{CR} = 23.74\%$ in LS-MBD.}\label{fig:proj-source-spec}
\vspace{-5mm}
\end{figure}


In LS-MBD, when $\text{CR} \geq 23.74\%$, the filter corresponding to $\proj$ is initialized at random. For lower ratios, we ``warm-start'' the network using a shortened version of the filter learned when $\text{CR} = 23.74\%$. Figure~\ref{fig:proj} shows, in the time domain, the compression filters learned for various compression ratios. Figure~\ref{fig:proj-source-spec} depicts the magnitude of the discrete Fourier transform of the source (black), learned compression filter (blue), and the compressed measurements (red) when $\text{CR}=23.74\%$. The alignment of $\z$ and $\s$ indicates that the filtering operation performed by the learned filter preserves information from the source, which may explain the success of LS-MBD compared to the other methods.

\vspace{-3mm}
\section{Conclusions}~\label{sec:con}
We proposed a compressive sparse multichannel blind-deconvolution method, named LS-MBD, based on unfolding neural networks~\cite{monga2019algorithm}. LS-MBD is an autoencoder that recovers sparse filters at the output of its encoder, and whose convolutional decoder corresponds to the source of interest. In this framework, we learn an efficient and structured compression matrix that allows us to have a faster and better accuracy in sparse filter recovery than other methods. We attribute our framework's superiority against FS-MBD~\cite{mulleti_mbd} to learning a compression optimized for both reconstruction and filter recovery.


\bibliographystyle{IEEEbib}
\bibliography{strings,icassp}

\end{document}